\documentclass{article}
\usepackage{spconf,amsmath,graphicx}
\usepackage{booktabs}
\usepackage[table]{xcolor}
\usepackage{url}
\usepackage{lmodern}
\newcommand{\tablecaptiongap}{\vspace{0.5em}}

\title{SURE-Voice: A Front-End Baseline for Speech-Evidence Filtering in Speech LLMs}
\name{Mengzhe Geng}
\address{National Research Council Canada\\{\tt Mengzhe.Geng@nrc-cnrc.gc.ca}}

\begin{document}
\maketitle
\pagestyle{plain}
\thispagestyle{plain}

\begin{abstract}
Speech language models (speech LLMs) can generate plausible outputs from audio that contains no usable speech evidence. We study this failure as a pre-generation support-estimation problem and present SURE-Voice, a training-free front end that decides whether an audio prompt contains intelligible speech evidence before calling a speech LLM. We build SURE-Challenge with a 640-example SURE-Core split and a 1,920-example SURE-Extended split derived from 120 LibriSpeech source utterances. Using one fixed operating point, an energy screen plus Whisper token confidence raises unsupported accuracy on the held-out Extended test from 0.000--0.133 to 0.919 for six non-degenerate speech LLM backbones, while supported accuracy remains 0.919--0.970 and downstream calls fall from 480 to 287. A 500-clip ESC-50 sanity set shows the same pattern on real environmental audio, with vocal non-speech as a residual failure mode. An overlap diagnostic shows that source attribution remains separate from speech-evidence filtering. The evidence supports a controlled benchmark baseline and a deployment-oriented analysis; it does not establish universal robustness to semantic answerability, gain variation or natural conversations.
\end{abstract}


\section{Introduction}

Speech-capable large language models can transcribe audio, follow spoken instructions and answer audio questions~\cite{qwenaudio,qwen2audio,qwen25omni}. They can also produce plausible text when a clip contains silence, environmental noise, synthetic tones, music-like events or several overlapping speakers. This behavior is costly in applications that need an answer grounded in the input: the system may answer before checking whether the requested evidence is present.

We study this failure as \emph{speech support estimation}. Given audio $x$ and a prompt $q$, a front end estimates whether $x$ contains enough intelligible speech evidence for the requested response. The downstream speech LLM is called only when the estimate passes a fixed gate; otherwise the system abstains or asks for clarification. This~task addresses a different question from semantic answerability: an acoustic gate cannot determine whether a question is answerable from a transcript, and speech confidence alone cannot identify a requested speaker in an overlap.

\begin{figure}[t]
\centering
\includegraphics[width=\columnwidth]{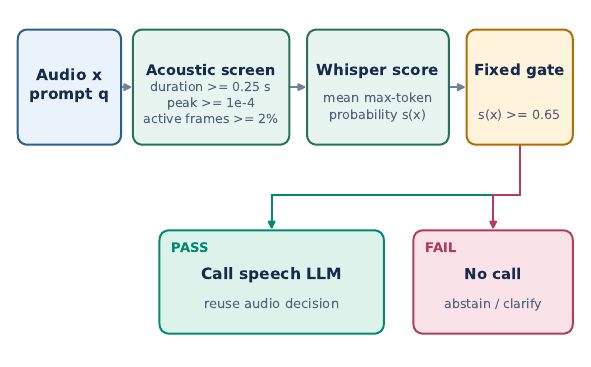}
\caption{SURE-Voice filters speech evidence before generation. The score $s(x)$ is the average maximum Whisper token probability; the fixed gate passes when $s(x)\geq 0.65$. Solid arrows show the fixed inference path, whose audio decision is reused across downstream backbones. The gate does not estimate semantic answerability or reliable target-speaker attribution.}
\label{fig:sure-flow}
\end{figure}

Figure~\ref{fig:sure-flow} shows the operating sequence. The front end is intentionally small and interpretable so that its effect can be separated from the downstream model. We ask three questions: whether unsupported-audio failures persist across backbones, whether a fixed pre-generation gate improves the supported/unsupported trade-off, and which residual errors require source attribution as a separate problem from speech detection.

Our contributions are threefold. First, we construct SURE-Challenge, a source-disjoint benchmark of supported and unsupported audio with explicit expected behavior labels at two scales. Second, we evaluate a training-free gate combining transparent acoustic checks with Whisper confidence and transfer its fixed decisions across six speech LLM backbones. Third, we separate non-speech rejection from source attribution through ESC-50 and an overlap diagnostic, which exposes the boundary of the proposed baseline.

\section{Related Work}

Large audio-language models such as Qwen-Audio, Qwen2-Audio, Qwen2.5-Omni, Audio Flamingo Next, Audio Flamingo 3, MiniCPM-o and Ultravox combine acoustic encoders with language-model decoders for speech and general audio understanding~\cite{qwenaudio,qwen2audio,qwen25omni,audioflamingonext,audioflamingo3,minicpmo,ultravox}. Audio hallucination benchmarks test unsupported contents over speech, environmental sound and music inputs~\cite{ahabench,halluaudio}, and Whisper is known to generate repetitive text on non-speech audio~\cite{whisper_nonspeech}. These studies provide broad diagnostics. SURE-Challenge isolates the earlier decision of whether the input contains speech evidence for the requested interaction; it is a support-estimation baseline, not a replacement for broad hallucination evaluation.

Automatic speech recognition (ASR) confidence estimation and voice activity detection (VAD) provide classical mechanisms for rejecting uncertain hypotheses and non-speech segments~\cite{asr_confidence_survey,vad_sohn1999}. Selective classification studies the related option of abstaining under uncertainty~\cite{selective_classification,selective_prediction_vlm}. Unsupported speech-LLM inputs also include intelligible but source-ambiguous speech. SURE-Voice therefore combines acoustic checks with Whisper token confidence~\cite{whisper} and treats source-count uncertainty as a separate residual problem~\cite{pyannote}.

\section{SURE-Challenge}

Each example contains an audio file, a user prompt, an expected behavior label and, for answerable cases, a reference transcript or answer. SURE-Core is derived from LibriSpeech validation-clean~\cite{librispeech}. Forty source utterances are selected with a seeded streaming shuffle while preferring distinct speakers. Source order defines disjoint splits with 20 train, 10 development and 10 held-out test source speakers. Each source contributes 16 derived examples, giving 640 examples and 320/160/160 train/development/test examples. SURE-Extended applies the same generator to LibriSpeech train-clean-100 and uses 120 source utterances and 1,920 examples with 960/480/480 splits. Its held-out test contains 270 supported and 210 unsupported examples. Table~\ref{tab:protocol} gives the complete composition used in the reported experiments.

\begin{table}[t]
\centering
\caption{SURE benchmark composition. Counts are examples per split. Babble is labeled abstain-or-clarify because the prompt asks for a main speaker without specifying a reliable source.}
\label{tab:protocol}
\tablecaptiongap
\scriptsize
\setlength{\tabcolsep}{2.4pt}
\begin{tabular}{lrrrrrr}
\toprule
& \multicolumn{3}{c}{\textbf{SURE-Core}} & \multicolumn{3}{c}{\textbf{SURE-Extended}} \\
\cmidrule(lr){2-4}\cmidrule(lr){5-7}
\textbf{Family} & \textbf{Tr.} & \textbf{Dev} & \textbf{Test} & \textbf{Tr.} & \textbf{Dev} & \textbf{Test} \\
\midrule
Answerable speech/QA & 180 & 90 & 90 & 540 & 270 & 270 \\
Silence/noise/tone & 100 & 50 & 50 & 300 & 150 & 150 \\
Babble abstain/clarify & 40 & 20 & 20 & 120 & 60 & 60 \\
\bottomrule
\end{tabular}
\end{table}

Supported examples include clean transcription, additive white-noise speech at 20/10/5 dB signal-to-noise ratio (SNR), 3.8 kHz low-pass filtering, 70 ms echo reverb with decay 0.35, 0.9/1.1 speed perturbation and first-word question answering. Unsupported non-speech examples use length-matched silence, white/pink/brown noise and three-tone synthetic music. Unsupported babble mixes two or four different source utterances with random 0.92--1.08 speed perturbations and 50--900 ms circular shifts, using the prompt \texttt{Transcribe the main speaker.} This makes babble a source-attribution test in addition to an ASR robustness test.

For answerable transcription, an output is correct when normalized word error rate (WER) is no larger than 0.25; first-word question answering uses normalized exact containment. The fixed scorer lowercases text, normalizes punctuation and whitespace, and removes conversational transcript prefixes before WER. Prefix removal is limited to the fixed scorer list recorded with the prediction carriers. Unsupported examples are correct only when the output contains a predefined abstention or clarification cue, including \texttt{abstain}, \texttt{unable to transcribe}, \texttt{no speech}, \texttt{no audio}, \texttt{insufficient audio}, \texttt{cannot determine}, \texttt{unclear}, \texttt{ambiguous}, \texttt{overlapping} or \texttt{multiple speakers}. Cue matching applies the same lowercase, punctuation and whitespace normalization. Bracketed event tags such as \texttt{[Music]} do not satisfy a transcript prompt. Abstaining on supported speech is an error. We report supported accuracy, unsupported accuracy, mean per-example WER over supported examples that pass the gate and receive a downstream prediction, and downstream call count. A gate-abstained supported example is excluded from WER but remains an error in supported accuracy. We also construct a 500-clip ESC-50~\cite{esc50} sanity set with ten clips from each of the 50 classes; it is not used for threshold selection or model development.

\section{Support Estimation}

Before calling the speech LLM, SURE-Voice makes a binary answer-or-abstain decision. The reported acoustic screen rejects clips with duration below 0.25 s, peak amplitude below $10^{-4}$ or non-silent frame ratio below 0.02. Mean root-mean-square (RMS) level is recorded in the prediction carriers for audit but is not part of the reported decision rule. Replaying every stored v5/v6 test carrier with the RMS condition removed produces the same headline decisions, because no stored decision depends on RMS alone. This is a compatibility audit of the evaluated carriers, not an external gain-robustness experiment.

Audio that passes the screen is decoded by Whisper-small with English transcription settings. Let $p_t$ be the maximum token probability at decoding step $t$. We use $s(x)=T^{-1}\sum_t p_t$ and reject when $s(x)<\tau$. The headline operating point is a fixed $\tau=0.65$, selected before the reported SURE-Extended test replay by first requiring supported accuracy of at least 0.90 on the Core development subset, then choosing the conservative point for the stored replay. It is transferred unchanged to all downstream backbones and the ESC-50 sanity set. On the 90 supported examples in that development sweep, it retains 89 (0.989). Retrospective Extended replay at higher thresholds improves unsupported accuracy but reduces supported coverage. The saved threshold sweeps are reported as retrospective sensitivity analyses; they do not redefine the headline test point.

For examples that pass the gate, the original audio and prompt are sent to six retained transcript backbones: Qwen2, Qwen2.5-Omni, Qwen-Audio-Chat, Audio Flamingo Next, Audio Flamingo 3 and MiniCPM-o 2.6. Core ablations use the three representative Qwen-family systems. Transfer reuses the same audio decisions and changes only downstream predictions, so gate selection uses no speech-LLM output. For a secondary source-attribution diagnostic, we train audio-only overlap classifiers using RMS, spectral, zero-crossing and mel-frequency cepstral coefficient (MFCC) summary features. Their thresholds are selected on development speakers under a supported-reject constraint of at most 0.05 and then frozen for test.

\section{Experiments}

\subsection{Held-Out Results}

We use SURE-Core for ablations, prompt analysis and threshold sensitivity, and SURE-Extended for the scaled held-out evaluation. Table~\ref{tab:core} reports the Core test results with Wilson 95\% intervals. The raw systems cover three representative backbones. Direct self-abstention uses the exact instruction: \texttt{If the audio contains no clear intelligible speech, only non-speech/noise/music/tone, or multiple overlapping speakers with no single attributable main speaker, respond exactly ABSTAIN.} Qwen2-Audio and Qwen2.5-Omni over-abstain on supported speech, while Qwen-Audio-Chat largely ignores the instruction.

\begin{table*}[t]
\centering
\caption{Held-out SURE-Core test results: 90 supported and 70 unsupported examples. Brackets give Wilson 95\% intervals; WER is reported as a percentage. Self rows use the same backbone with an explicit abstention instruction. Integrated rows use the model-independent fixed $\tau=0.65$ gate.}
\label{tab:core}
\tablecaptiongap
\scriptsize
\setlength{\tabcolsep}{3.5pt}
\begin{tabular}{lrrrr}
\toprule
\textbf{System} & \textbf{Supported acc. $\uparrow$} & \textbf{Unsupported acc. $\uparrow$} & \textbf{WER (\%) $\downarrow$} & \textbf{Calls $\downarrow$} \\
\midrule
Raw Qwen2-Audio & 0.844 [0.756, 0.905] & 0.014 [0.003, 0.077] & 13.6 & 160 \\
Self-abstain prompt + Qwen2-Audio & 0.033 [0.011, 0.094] & 1.000 [0.948, 1.000] & 34.2 & 160 \\
Energy gate + Qwen2-Audio & 0.844 [0.756, 0.905] & 0.157 [0.090, 0.260] & 13.6 & 150 \\
Energy+spectral gate + Qwen2-Audio & 0.578 [0.475, 0.675] & 0.471 [0.359, 0.587] & 13.0 & 98 \\
\rowcolor{gray!15}
Integrated support gate + Qwen2-Audio & 0.833 [0.743, 0.896] & 0.914 [0.825, 0.960] & 13.7 & 96 \\
Raw Qwen-Audio-Chat & 0.744 [0.646, 0.823] & 0.000 [0.000, 0.052] & 22.0 & 160 \\
Self-abstain prompt + Qwen-Audio-Chat & 0.300 [0.215, 0.401] & 0.014 [0.003, 0.077] & 70.3 & 160 \\
\rowcolor{gray!15}
Integrated support gate + Qwen-Audio-Chat & 0.744 [0.646, 0.823] & 0.914 [0.825, 0.960] & 22.0 & 96 \\
Raw Qwen2.5-Omni & 0.944 [0.877, 0.976] & 0.000 [0.000, 0.052] & 5.5 & 160 \\
Self-abstain prompt + Qwen2.5-Omni & 0.156 [0.095, 0.244] & 1.000 [0.948, 1.000] & 0.0 & 160 \\
\rowcolor{gray!15}
Integrated support gate + Qwen2.5-Omni & 0.944 [0.877, 0.976] & 0.914 [0.825, 0.960] & 5.5 & 96 \\
\bottomrule
\end{tabular}
\end{table*}

The integrated gate raises unsupported Qwen2-Audio accuracy from 0.014 to 0.914 on SURE-Core, while supported accuracy is 0.833 and calls fall from 160 to 96 (40\%). The acoustic controls show why the confidence stage is useful: energy-only rejection is conservative, whereas the energy-plus-spectral rule rejects too much supported speech. The self-abstention results show that an instruction alone does not provide a stable support decision.

\begin{table}[t]
\centering
\caption{Scaled held-out SURE-Extended test results: 270 supported and 210 unsupported examples from held-out source utterances. Brackets give Wilson 95\% intervals; S. and U. denote supported and unsupported accuracy, and WER is reported as a percentage. SURE rows replay the same audio-only fixed $\tau=0.65$ gate across six retained transcript backbones.}
\label{tab:extended}
\tablecaptiongap
\scriptsize
\setlength{\tabcolsep}{0.6pt}
\begin{tabular}{@{}l@{\hspace{1.5pt}}r@{\hspace{1.5pt}}r@{\hspace{1.5pt}}r@{\hspace{1.5pt}}r@{}}
\toprule
\textbf{System} & \textbf{S. $\uparrow$} & \textbf{U. $\uparrow$} & \textbf{WER (\%) $\downarrow$} & \textbf{Calls $\downarrow$} \\
\midrule
Raw Qwen2 & .930 [.893,.954] & .071 [.044,.114] & 4.3 & 480 \\
\rowcolor{gray!15}
SURE+Qwen2 & .930 [.893,.954] & .919 [.874,.949] & 4.3 & 287 \\
Raw Omni & .967 [.938,.982] & .000 [.000,.018] & 3.5 & 480 \\
\rowcolor{gray!15}
SURE+Omni & .967 [.938,.982] & .919 [.874,.949] & 3.5 & 287 \\
Raw Chat & .919 [.880,.946] & .000 [.000,.018] & 20.3 & 480 \\
\rowcolor{gray!15}
SURE+Chat & .919 [.880,.946] & .919 [.874,.949] & 20.3 & 287 \\
Raw AF-Next & .956 [.924,.974] & .000 [.000,.018] & 4.5 & 480 \\
\rowcolor{gray!15}
SURE+AF-Next & .956 [.924,.974] & .919 [.874,.949] & 4.5 & 287 \\
Raw AF3 & .970 [.943,.985] & .000 [.000,.018] & 2.9 & 480 \\
\rowcolor{gray!15}
SURE+AF3 & .970 [.943,.985] & .919 [.874,.949] & 2.9 & 287 \\
Raw MiniCPM-o & .948 [.915,.969] & .133 [.094,.186] & 4.0 & 480 \\
\rowcolor{gray!15}
SURE+MiniCPM-o & .948 [.915,.969] & .919 [.874,.949] & 4.0 & 287 \\
\bottomrule
\end{tabular}
\end{table}

Table~\ref{tab:extended} shows that the fixed gate raises unsupported accuracy to 0.919 for every listed backbone while preserving each stored supported accuracy. For Qwen2-Audio, calls fall from 480 to 287, a 40\% reduction. The gate is audio-only and is not recalibrated for the six downstream systems. The attempted Ultravox v0.3 run is excluded from the headline table because it produced degenerate repeated punctuation under the deterministic transcript prompt.

Threshold sensitivity exposes the operating-point trade-off. On the stored Extended Qwen2 replay, $\tau=0.65$ gives unsupported accuracy 0.919 and 287 calls. The retrospective points $\tau=0.70,0.75,0.80$ give 0.962, 0.976, 0.986 and 278, 272, 262 calls. Supported accuracy at those points is 0.930, 0.919 and 0.889. The paper retains the fixed conservative $\tau=0.65$ for the headline comparison. On SURE-Core, a paired exact McNemar check gives 63 unsupported examples where SURE+Qwen2-Audio is correct and raw Qwen2-Audio is incorrect, versus 0 in the opposite direction.

A real-audio sanity check uses 500 ESC-50 clips, ten from each class. Under the same transcription prompt, the six headline backbones plus the attempted Ultravox baseline obtain 0.000--0.052 unsupported accuracy, whereas replaying the fixed gate raises the range to 0.840--0.854 and reduces calls from 500 to 80. Errors concentrate in human vocal or quasi-speech classes such as crying baby, laughing, breathing and clapping. This is a fixed sanity check, not a broad real-world benchmark; it shows that unsupported-audio failures are not limited to the synthetic silence, noise and tone families.

\subsection{Failure Analysis}

Table~\ref{tab:family} breaks down SURE+Qwen2-Audio on the Extended test. The non-speech families are separated from babble so the main support-estimation result is not conflated with source attribution.

\begin{table}[t]
\centering
\caption{SURE+Qwen2-Audio family breakdown on SURE-Extended test at the fixed $\tau=0.65$ gate.}
\label{tab:family}
\tablecaptiongap
\begin{tabular}{lrr}
\toprule
\textbf{Family} & \textbf{Success/total} & \textbf{Accuracy $\uparrow$} \\
\midrule
Clean speech & 28/30 & 0.933 \\
Speech+noise & 83/90 & 0.922 \\
Low-pass/reverb/speed & 112/120 & 0.933 \\
Speech QA & 28/30 & 0.933 \\
Silence/noise/tone & 150/150 & 1.000 \\
Overlap babble & 43/60 & 0.717 \\
\bottomrule
\end{tabular}
\end{table}

Table~\ref{tab:family} shows 150/150 correct rejections for silence, noise and tone. Raw Qwen2-Audio errors include Chinese text for silence, \texttt{Kids are talking by the door} for white or brown noise, and Chinese text for synthetic tones; the energy screen or low Whisper confidence removes these cases. Babble remains difficult because speech is present but the requested main speaker is not reliably attributable. The logistic audio-only overlap gate reaches development/test receiver operating characteristic area under the curve (ROC-AUC) of 0.972/0.966. Its frozen test threshold catches 50/60 overlaps but rejects 10/270 supported examples; this separate gate increases unsupported accuracy from 0.919 to 0.981 while supported accuracy decreases from 0.930 to 0.882. This is a source-attribution diagnostic and a coverage--safety trade-off, not a complete diarization solution.

\section{Discussion}

Supported-speech recognition and unsupported-audio robustness are different objectives. The raw backbones provide strong supported performance but still answer many unsupported inputs. Direct prompting is unstable, and transparent acoustic controls either miss energetic non-speech audio or reject noisy supported speech. SURE-Voice operates before generation and transfers the same audio decision across downstream models, which explains the stable unsupported-accuracy gain in Table~\ref{tab:extended}.

The evaluation scope is essential. SURE-Challenge uses controlled LibriSpeech-derived perturbations and synthetic overlap; ESC-50 supplies a small real environmental and vocal non-speech sanity set but does not represent natural conversations. The gate estimates acoustic support, not semantic answerability, fairness across speakers, gain robustness or reliable target-speaker identity. Natural meetings, far-field recordings, multilingual code-switching, vocal music and diarization errors require additional evaluation. The overlap result identifies this source-attribution problem without solving it.

For reproducibility, the local evidence is organized around the SURE-Core and SURE-Extended JSONL manifests, the fixed prompts and scorer, and stored raw/integrated prediction carriers named in the accompanying project README. The paper source package contains this manuscript, bibliography, ICASSP style and bibliography-style files, plus the revised figure source and export; it excludes unavailable external-audit or stress-test artifacts. A release should add the manifest generator, scorer, feature-replay script and checksums for each prediction carrier before the benchmark is used as a public comparison.

\section{Conclusion}

We presented SURE-Voice, a training-free front end for speech-evidence filtering before speech-LLM generation. On the source-disjoint SURE-Challenge evaluation, raw Qwen-Audio-Chat, Qwen2-Audio, Qwen2.5-Omni, Audio Flamingo Next, Audio Flamingo 3 and MiniCPM-o 2.6 show strong supported-speech performance but unreliable rejection of unsupported inputs. A fixed energy and Whisper-confidence gate raises unsupported accuracy to 0.919 for all six stored non-degenerate transcript backbones while retaining their supported accuracies and reducing downstream calls. ESC-50 confirms the pattern on a small real-audio sanity set, and overlap analysis shows that source attribution remains a separate bottleneck. These results support speech-evidence filtering as a useful front-end baseline within the tested controlled scope.

\bibliographystyle{IEEEbib}
\small
\bibliography{refs}

@article{qwenaudio,
  title={Qwen-Audio: Advancing Universal Audio Understanding via Unified Large-Scale Audio-Language Models},
  author={Chu, Yunfei and Xu, Jin and Zhou, Xiaohuan and Yang, Qian and Zhang, Shiliang and Yan, Zhijie and Zhou, Chang and Zhou, Jingren},
  journal={arXiv preprint arXiv:2311.07919},
  year={2023},
  doi={10.48550/arXiv.2311.07919}
}

@article{qwen2audio,
  title={Qwen2-Audio Technical Report},
  author={Chu, Yunfei and Xu, Jin and Yang, Qian and Wei, Haojie and Wei, Xipin and Guo, Zhifang and Leng, Yichong and Lv, Yuanjun and He, Jinzheng and Lin, Junyang and Zhou, Chang and Zhou, Jingren},
  journal={arXiv preprint arXiv:2407.10759},
  year={2024},
  doi={10.48550/arXiv.2407.10759}
}

@article{qwen25omni,
  title={Qwen2.5-Omni Technical Report},
  author={{Qwen Team}},
  journal={arXiv preprint arXiv:2503.20215},
  year={2025},
  doi={10.48550/arXiv.2503.20215}
}

@misc{audioflamingo3,
  title={{Audio Flamingo 3}: Advancing Audio Intelligence with Fully Open Large Audio Language Models},
  author={{NVIDIA}},
  howpublished={Model card},
  year={2026},
  url={https://huggingface.co/nvidia/audio-flamingo-3-hf}
}

@misc{minicpmo,
  title={{MiniCPM-o 2.6}: A {GPT-4o}-Level {MLLM} for Vision, Speech and Multimodal Live Streaming on Your Phone},
  author={{OpenBMB Team}},
  howpublished={Technical report and model card},
  year={2025},
  url={https://huggingface.co/openbmb/MiniCPM-o-2_6}
}

@inproceedings{librispeech,
  title={{LibriSpeech}: An {ASR} Corpus Based on Public Domain Audio Books},
  author={Panayotov, Vassil and Chen, Guoguo and Povey, Daniel and Khudanpur, Sanjeev},
  booktitle={Proceedings of the IEEE International Conference on Acoustics, Speech and Signal Processing},
  pages={5206--5210},
  year={2015},
  doi={10.1109/ICASSP.2015.7178964}
}

@inproceedings{whisper,
  title={Robust Speech Recognition via Large-Scale Weak Supervision},
  author={Radford, Alec and Kim, Jong Wook and Xu, Tao and Brockman, Greg and McLeavey, Christine and Sutskever, Ilya},
  booktitle={Proceedings of the 40th International Conference on Machine Learning},
  pages={28492--28518},
  year={2023}
}

@inproceedings{pyannote,
  title={pyannote.audio: Neural Building Blocks for Speaker Diarization},
  author={Bredin, Herv{\'e} and Yin, Ruiqing and Coria, Juan Manuel and Gelly, Gregory and Korshunov, Pavel and Lavechin, Marvin and Fustes, Diego and Titeux, Hadrien and Bouaziz, Wassim and Gill, Marie-Philippe},
  booktitle={Proceedings of the IEEE International Conference on Acoustics, Speech and Signal Processing},
  pages={7124--7128},
  year={2020},
  doi={10.1109/ICASSP40776.2020.9052974}
}

@inproceedings{halluaudio,
  title={{HalluAudio}: A Comprehensive Benchmark for Hallucination Detection in Large Audio-Language Models},
  author={Zhao, Feiyu and Chen, Yiming and Lu, Wenhuan and Zhang, Daipeng and Yue, Xianghu and Wei, Jianguo},
  booktitle={Proceedings of the 64th Annual Meeting of the Association for Computational Linguistics},
  year={2026},
  doi={10.48550/arXiv.2604.19300}
}

@inproceedings{ahabench,
  title={{AHa-Bench}: Benchmarking Audio Hallucinations in Large Audio-Language Models},
  author={Cheng, Xize and Fu, Dongjie and Wen, Chenyuhao and Yu, Shannon and Wang, Zehan and Ji, Shengpeng and Arora, Siddhant and Jin, Tao and Watanabe, Shinji and Zhao, Zhou},
  booktitle={Advances in Neural Information Processing Systems, Datasets and Benchmarks Track},
  year={2025}
}

@article{whisper_nonspeech,
  title={Investigation of Whisper {ASR} Hallucinations Induced by Non-Speech Audio},
  author={Bara{\'n}ski, Mateusz and Jasi{\'n}ski, Jan and Bartolewska, Julitta and Kacprzak, Stanis{\l}aw and Witkowski, Marcin and Kowalczyk, Konrad},
  journal={arXiv preprint arXiv:2501.11378},
  year={2025},
  doi={10.48550/arXiv.2501.11378}
}

@inproceedings{selective_classification,
  title={Selective Classification for Deep Neural Networks},
  author={Geifman, Yonatan and El-Yaniv, Ran},
  booktitle={Advances in Neural Information Processing Systems},
  year={2017},
  pages={4878--4887}
}

@inproceedings{selective_prediction_vlm,
  title={Selective {``}Selective Prediction{''}: Reducing Unnecessary Abstention in Vision-Language Reasoning},
  author={Srinivasan, Tejas and Hessel, Jack and Gupta, Tanmay and Lin, Bill Yuchen and Choi, Yejin and Thomason, Jesse and Chandu, Khyathi Raghavi},
  booktitle={Findings of the Association for Computational Linguistics: ACL 2024},
  year={2024}
}

@inproceedings{esc50,
  title={{ESC}: Dataset for Environmental Sound Classification},
  author={Piczak, Karol J.},
  booktitle={Proceedings of the 23rd ACM International Conference on Multimedia},
  pages={1015--1018},
  year={2015},
  doi={10.1145/2733373.2806390}
}

@article{asr_confidence_survey,
  title={Confidence Measures for Speech Recognition: A Survey},
  author={Jiang, Hui},
  journal={Speech Communication},
  volume={45},
  number={4},
  pages={455--470},
  year={2005},
  doi={10.1016/j.specom.2004.12.004}
}

@article{vad_sohn1999,
  title={A Statistical Model-Based Voice Activity Detection},
  author={Sohn, Jongseo and Kim, Nam Soo and Sung, Wonyong},
  journal={IEEE Signal Processing Letters},
  volume={6},
  number={1},
  pages={1--3},
  year={1999},
  doi={10.1109/97.736233}
}

@article{audioflamingonext,
  title={Audio Flamingo Next: Next-Generation Open Audio-Language Models for Speech, Sound, and Music},
  author={Ghosh, Sreyan and Goel, Arushi and Jayakumar, Kaousheik and Anand, Nishit and Kong, Zhifeng and Gururani, Siddharth and Lee, Sang-gil and Kim, Jaehyeon and Aljafari, Aya and Yang, Chao-Han Huck and Kim, Sungwon and Duraiswami, Ramani and Manocha, Dinesh and Shoeybi, Mohammad and Catanzaro, Bryan and Liu, Ming-Yu and Ping, Wei},
  journal={arXiv preprint arXiv:2604.10905},
  year={2026}
}

@misc{ultravox,
  title={Ultravox: A Fast Multimodal LLM for Real-Time Voice},
  author={{Fixie AI}},
  howpublished={\url{https://github.com/fixie-ai/ultravox}},
  year={2024}
}

\end{document}